# Cryogenic digital data links for the liquid argon time projection chamber

**Tiankuan Liu[a+*], Datao Gong[a], Suen Hou[b], Chonghan Liu[a], Da-Shung Su[b], Ping-kun Teng[b], Annie C. Xiang[a], and Jingbo Ye[a+]**

[a] *Southern Methodist University,
  Dallas, Texas 75275, USA*

[b] *Institute of Physics, Academia Sinica,
  Nangang 11529, Taipei, Taiwan*

[+] On behalf of the LBNE collaboration
  *E-mail*: tliu@mail.smu.edu

ABSTRACT: In this paper we present the cryogenic functionality of the components of data links for the Liquid Argon Time Projection Chamber (LArTPC), a potential far site detector technology of the Long-Baseline Neutrino Experiment (LBNE). We have confirmed that an LVDS driver can drive a 20-meter CAT5E twisted pair up to 1 gigabit per second at the liquid nitrogen temperature (77 K). We have verified that a commercial-off-the-shelf (COTS) serializer, a laser diode driver, laser diodes, optical fibers and connectors, and field-programming gate arrays (FPGA's) continue to function at 77 K. A variety of COTS resistors and capacitors have been tested at 77 K. All tests we have conducted show that the cryogenic digital data links for the liquid argon time projection chamber are promising.

# Contents

## 1. Introduction

### 1.1 LBNE and LArTPC

The Long-Baseline Neutrino Experiment (LBNE) is a proposed experiment to explore the interactions and transformations of a high-intensity neutrino beam by sending it from Fermi National Accelerator Laboratory (FNAL) 1300 kilometers through the earth to Homestake Mine in Lead, South Dakota [1]. The Liquid Argon Time Projection Chamber (LArTPC) has full 3-D event reconstruction capability with sub-millimeter position resolution, larger than 90% electron-photon separation, and particle energy threshold as low as 1-2 MeV, making it an ideal neutrino physics detector [2]. Two LArTPC modules each of which has about 20-kton volume have been proposed as a potential far-site detector technology of LBNE. The increase of the input capacitance and the input equivalent noise makes it impossible to build a large fiducial volume detector with all preamplifiers located outside the cryostat at room temperature (RT). Moreover, large amount of cables and feedthroughs increase cost, thermal load, the possibility of outgassing and leaks, and the failure rate. Therefore, a cold front-end electronics system has been chosen as the reference design of the LBNE LArTPC [3].

    In the cold electronics system, all preamplifiers, shapers, analog to digital converters (ADC's), zero suppression system, digital buffers, data multiplexers, and cable drivers are mounted in the liquid argon cryostat. To develop a cold electronics system, it is essential that all components still function properly at the liquid argon temperature (about 89 K). Since the cold electronics system has little or no access for repair or replacement after installation, the lifetime

of all components operating in the LArTPC must be more than the design lifetime of the LArTPC or 20 years.

The application-specific integrated circuits (ASIC's), including the preamplifiers, shapers, ADCs, zero suppression, and digital buffers, are under development and the ASIC performances at cryogenic temperatures are discussed elsewhere [3]. In this paper we will focus on the data links, including multiplexers and cable drivers.

### 1.2 Cold data link overview

Figure 1 is the block diagram of the cold electronics. The analog signal from a sense wire is integrated in a charge sensitive amplifier (CSA) and converted into a voltage signal. The voltage signal is filtered in a shaper and digitized into a digital signal in a 12-bit ADC sampling at 2 mega-samples per second. Following the ADC is the zero suppression and digital buffers. A 16-channel mixed signal front-end ASIC, including CSAs, shapers, ADCs, zero suppression, and buffers, is under development. The signals out of the eight mixed signal ASICs will be multiplexed in a 128-channel front-end mother board. The signals out of 30 front-end mother boards will be multiplexed and transmitted out of the cryostat in a mux/control board.

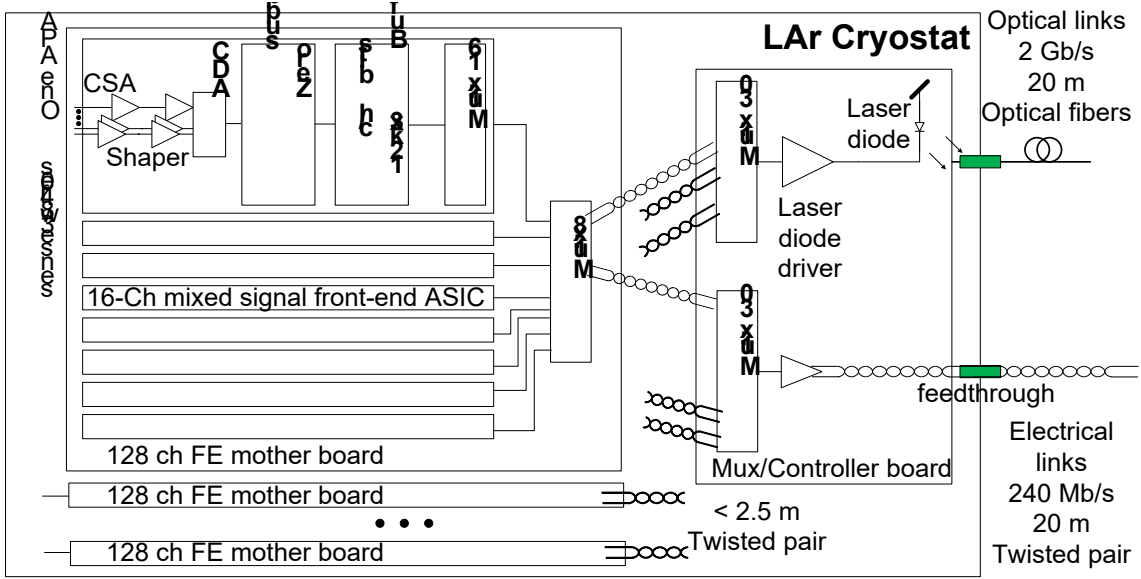

Figure 1: The block diagram of the cold front-end electronics

The data rate out of the LArTPC highly depends on where the LArTPC is located. The data rate of a LArTPC at 244-meter underground is about 240 Mbps per anode plane assembly (APA), including a safety factor of 20. In this situation, the data will be transmitted through electrical data links. If the LArTPC is located on the earth surface, the data rate will be as high as 2 gigabit per second (Gbps) per APA before considering any safety factor. In this situation, the data will be transmitted through optical data links. The distance of data links is no more than 20 meters for both electrical and optical links. Both the optical and electrical links are shown in Figure 1, but only one of the technologies will be used in the end. For optical links, the links between a mother board and a mux/control board are still electrical. For either electrical or optical link, double or triple links rather than a single link may be used to improve the reliability.

For electrical links, a natural choice is to transmit Low-voltage differential signaling (LVDS) over twisted pairs like Catalog 5 (CAT5) or enhanced Catalog 5 (CAT5E). Though LVDS and twisted pairs have been used for a long time and a lot of experience has been gained [4-6], it is not clear that 240 Mbps LVDS signals can be transmitted over 20 meters CAT5 or CAT5E twisted pairs at the liquid argon temperature.

An optical link is more complex than an electrical link. An optical link on the transmitter side includes a low-speed multiplexer, an encoder to provide a transmission control protocol, parallel word boundary in serial data, and DC balance, a serializer to convert parallel data to serial data (sometimes the encoder is included in a serializer), a laser diode driver, a laser diode, an optical fiber and an optical connector. The questions to be answered include whether the components of optical links still function at the liquid argon temperature.

**1.3 Challenges in cold electronics**

The data links can be implemented in either application-specific integrated circuits (ASIC's) or commercial-off-the-shelf (COTS) devices. In both approaches, the same two challenges exist: the functionality and the reliability.

The first challenge the cold electronics faces is the functionality. The functionality of all advanced electronic circuits relies on semiconductor transistors. At certain temperatures (about 40 K for silicon and 20 K for germanium) the dopants in a semiconductor transistor lack thermal energy to ionize and produce enough carriers [7]. Silicon bipolar junction transistors (BJT's) suffer such carrier freezing-out at cryogenic temperatures, but their performance actually degrades at higher temperatures. In fact, when the temperature is below about 100 K, silicon BJT's lose the gain rapidly and become unusable because of the low efficiency of emitter-base injection. Silicon enhancement MOSFET's (including CMOS) do not have the carrier freezing-out issue because the electric field from the gate can ionize the carriers. Heterojunction bipolar transistors (HBT's), including those based on SiGe, can operate down to very low cryogenic temperatures because their bandgap is much smaller.

Even if a semiconductor transistor can operate at very low temperatures, its parameters like threshold voltage and transconductance may change a lot when the temperature drops from RT to the liquid argon temperature. The issue of the parameter change is solved in quite different ways in ASIC and COTS approaches.

In the ASIC approach, a cryogenic device model may not be available in the design kit of most commercial processes. A cryogenic device model needs to be established before an ASIC is designed. There is no fundamental problem in the ASIC approach. However, it costs a lot of time and efforts to develop an ASIC.

In contrast, COTS devices are cheap and easy to acquire. The problem with COTS devices is that current COTS devices are rarely designed to operate at the liquid argon temperature. Our goal is to evaluate if the COTS devices we choose still perform properly at the liquid argon temperature.

The second challenge to operate a device at the liquid argon temperature is the reliability. Most electronic devices have longer lifetime with reduced temperatures. For example, a common failure mode in electronic devices is electromigration, the atom movement in the metal wires caused by an electromagnetic field. The mean time to failure (MTTF) of a copper wire due to electromigration is improved by over 37 orders of magnitude when the operating temperature changes from 300 K to 89 K, assuming a typical activity energy of 0.9 eV [8]. However, the operation at cryogenic temperatures may induce new failure modes. One known failure mode is hot carrier effects [8-11]. When the temperature decreases, channel carriers see less phonon scattering, so their mean free path lengths increase. As a result, more channel carriers (hot carriers) gain enough energy to break the Si–Si bonds and generate more impact ionization. These electrons are trapped in the gate oxide, causing transconductance degradation and threshold voltage shift. The failure caused by hot carrier effects is a major concern to operate an advanced MOSFET device at cryogenic temperatures.

For an ASIC, a common procedure is to study the lifetime of a transistor in the worst case and then apply some constraints in the design. For a COTS device, since we do not have any control on the intrinsic structure, the only thing we can do is either to verify that there is no performance degradation at cryogenic temperatures or to measure the device lifetime. Since the hot carrier effects are weaker at higher temperatures, it is impossible to accelerate the lifetime test by increasing the temperature, as is usually done in a lifetime test. By means of increasing the power voltage, the operating current, or the operating frequency, or by lowering the operating temperature, the test can be accelerated. However, the study on the lifetime at cryogenic temperatures is beyond the scope of this paper and will not be discussed further in this paper.

We are seeking the COTS approach first unless we see a showstopper. In most cases, we choose a device which meets our requirements, is produced in the main stream technology, and is easy to be acquired. If a device (for example, a laser diode) may be made in quite different ways, we test at least one sample in each type.

Since liquid nitrogen is much cheaper than liquid argon and the temperature difference between liquid nitrogen (77 K) and liquid argon (89 K) is small, in this paper all studies were conducted at the liquid nitrogen temperature or at the temperatures from RT to 77 K.

## 2. Cryogenic optical links

For optical links, we tried to answer whether the components of optical links still function at the liquid argon temperature. We have tested optical fibers and optical connectors, laser diodes, a transceiver, and a laser diode driver.

### 2.1 Optical fibers and optical connectors

A single mode (SM) optical fiber (Part number SMF-28 from Corning) and a multimode (MM) optical fiber (Part number InfiniCor SX+ from Corning), five SM LC connectors (Part number F1-8005 from Fiber Instrument Sales, Inc.) and five MM LC connectors (Part number F1-8001 Fiber Instrument Sales, Inc.) have been tested from RT to 77 K [12]. The SM fiber and connectors are tested with 1310-nm laser, while the MM fiber and connectors are tested with 850 nm laser. From RT to 77 K, the insertion loss of fibers changes 0.034 ± 0.015 dB/m and 0.005 ± 0.002 dB/m for MM and SM fibers, respectively, while the insertion loss of connectors changes 0.139 ± 0.020 dB per connector and -0.284 ± 0.014 dB per connector for MM and SM connectors, respectively. The insertion loss change of optical fibers at cryogenic temperatures is caused by mismatch between the thermal expansion coefficients of the fiber and the cabling materials [13]. The insertion loss change of optical connectors at cryogenic temperatures is probably due to the misalignment between fibers. Given a usual optical power budget of around 10 dB in a link system, the small power loss in the fibers and connectors at 77 K can be easily accommodated.

### 2.2 Laser diodes

The first generation of laser diodes, which were based on homojunction structure, operated only at cryogenic temperatures because the threshold current was too high at RT [14]. Since then intensive research efforts were made to improve the carrier confinement and the optical guiding in order to bring down the threshold current at RT. Today, most laser diodes used in data communications are not cooled to cryogenic temperatures. This is the key reason for laser diodes to become mass-produced devices.

There is no fundamental difficulty to design a laser diode operating properly at cryogenic temperatures. However, considering the cost and the development time of a custom design, we decide to evaluate some COTS laser diodes to see if they can operate at cryogenic temperatures.

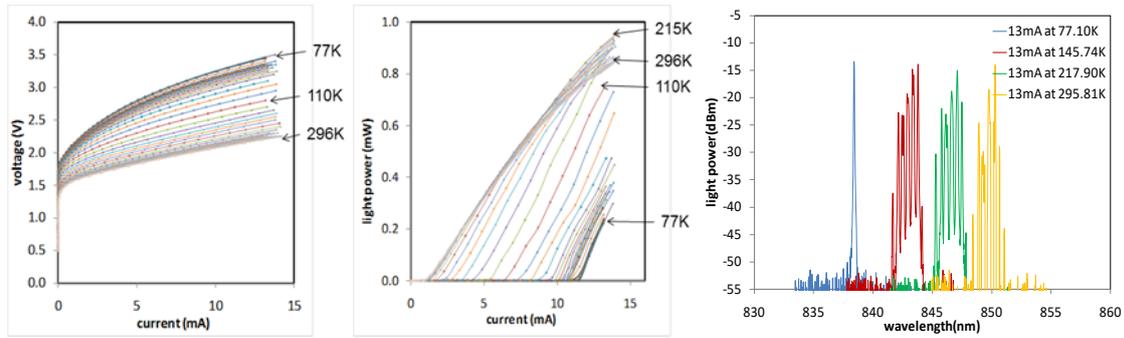

Figure 2: (from left to right) I-V curves (a), L-I curves (b) and optical spectra (c) of the VCSEL diode

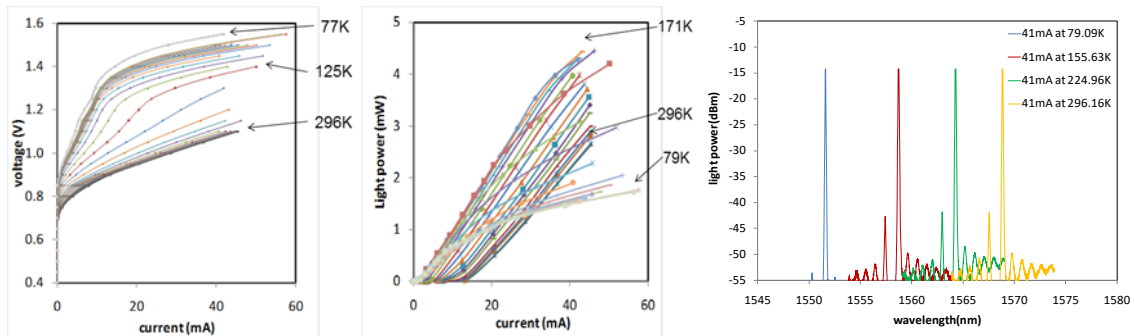

Figure 3: (from left to right) I-V curves (a), L-I curves (b) and optical spectra (c) of the DFB laser diode

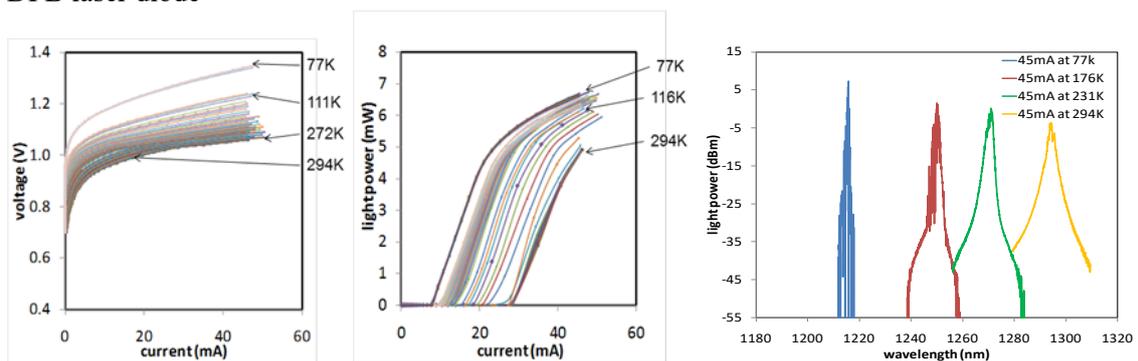

Figure 4: (from left to right) I-V curves (a), L-I curves (b) and optical spectra (c) of the FP laser diode

A vertical cavity surface emitting laser (VCSEL) diode, a distributed feedback (DFB) laser diode, and a Fabry-Pérot (FP) laser diode were tested from RT to 77 K [12]. The light power-current-voltage (L-I-V) characteristic curves and optical spectrum of each laser diode were measured in the test and displayed in Figures 2-4. Although the electrical and optical operating conditions of these devices showed slight or moderate changes, all three devices remained to be functional from RT to 77 K.

The threshold voltages of all three laser diodes increase from RT to 77 K. This is expected as the bandgap energy of active region increases [15]. The threshold currents of the DFB and FP laser diodes decrease when the temperature decreases from RT to 77 K, whereas the threshold current of the VCSEL increases. The threshold current changes of these three devices also conform to conventional observations [16]. The temperature dependence of the threshold currents of FP and DFB devices is usually expressed in exponential functions with a characteristic temperature [17]. As temperatures decrease, spontaneous emissions decrease, thus the threshold currents also decrease. In the case of VCSEL, the mismatch between the cavity wavelength selection and the gain profile brings about an increased threshold current as the temperature reduces.

The increment of the threshold voltage of all three devices and the threshold current of the VCSEL will increase the power consumption and potentially raise the concern of reliability. However, as is discussed in Section 1.3, all reliabilities which can be modelled in Arrhenius Equation are improved by many orders of magnitude with the cryogenic operation. Therefore, the increment of the power consumption is probably not an issue.

The light efficiencies of the VCSEL and the FP laser are almost unchanged at RT and 77 K, but the light efficiency of the DFB laser drops when the temperature is close to 77 K. The measurements include the effects from both the internal quantum efficiency and the external fiber-coupling efficiency. In Section 2.1, we have shown that the change of the fiber-coupling efficiency in an optical connector is small, but the fiber-coupling efficiency may differ from assembly to assembly.

In optical spectra, when the temperature changes from RT to 77 K, the central wavelengths of all three devices move toward the short wavelengths. The wavelength shift reflects the bandgap energy increase at reduced temperatures. The lasing wavelengths of the DFB and the VCSEL are determined by both the distributed mirror mode selection and the gain profile. The temperature dependence of the refractive index is generally small. Thus the changes of the central wavelengths of the DFB and the VCSEL are smaller.

The wavelength shift results in the sensitivity changes of photodiodes and the attenuation increase of fibers, but neither effect is significant. For example, the attenuation change due to the maximum wavelength shift in the FP laser diode in a 20-meter fiber is only about 0.001 dB.

In optical spectra at certain currents, when the temperature changes from RT to 77 K, the number of lasing modes of the VCSEL decreases and the spectral width becomes narrower. This may be only true for the tested VCSEL because lasing modes depend on the gain profile and mirror reflectivity. The spectral width of the FP laser decreases while the linewidth of the DFB laser changes very little. As a rule of thumb, the narrower spectral width improves the link power budget [18].

**2.3 The serializer**

We have tested a COTS serializer TLK2501 from Texas Instruments, Inc. The device was directly dipped into liquid nitrogen. We used a $2^7$-1 pseudorandom binary sequence (PRBS) generator embedded inside TLK2501 and measured eye diagrams and bit error rate (BER). The eye diagrams of the serializer at RT and at 77 K are shown in Figure 5. The data rate shown in the figure is 2.5 Gbps. The serializer has wider eye opening, faster rise time and fall time, smaller jitter and larger amplitude at 77 K than at RT, meaning the serializer worked even better at 77 K than at RT. The jitter and amplitude at 77 K and at RT fall within the specifications. The BERs at RT and at 77 K are $3.5\times10^{-13}$ and $2.8\times10^{-13}$, respectively at 90% confidence level.

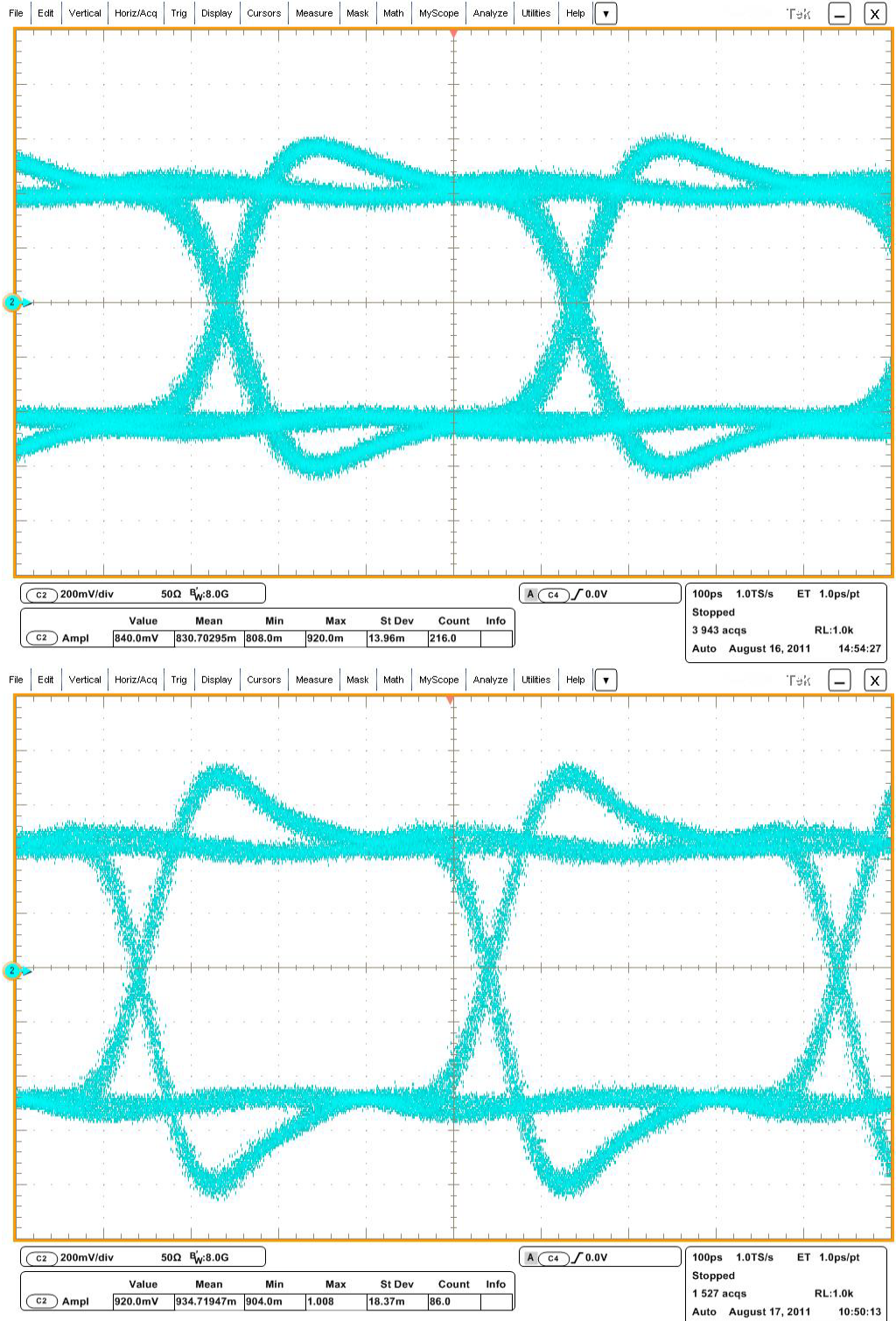

Figure 5: (from top to bottom) eye diagrams of TLK2501 at RT (a) and 77 K (b)

## 2.4 The laser diode driver

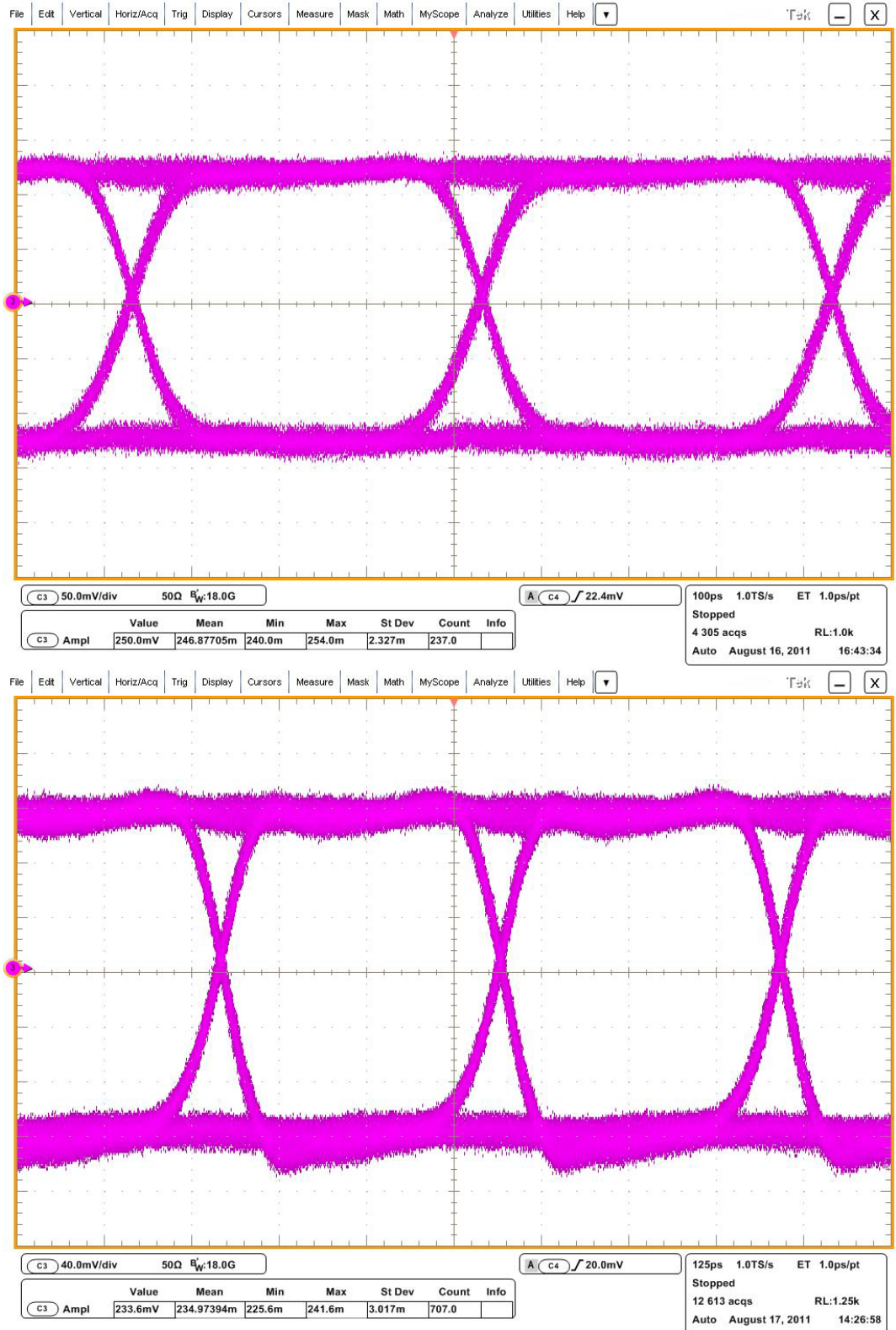

Figure 6: (from top to bottom) eye diagrams of MAX3850 at RT and 3.3 V (a) and at 77 K and 3.6 V (b)

MAX3850 from Maxim Integrated Products, Inc. is a laser diode driver fabricated in a silicon germanium bipolar process for data communication applications up to 2.7 Gbps. The device can drive both VCSEL's and edge-emitting laser diodes, either DC coupling or AC coupling with a single voltage power supply.

We tested MAX3850 at 77 K with a 15-Ω equivalent load, which is optimal for MAX3850 and close to the equivalent series resistance of an edge emitting laser diode. The device was directly dipped into liquid nitrogen and eye diagrams were measured. We did not measure the BER because the amplitude over the 15 Ω load is too small for a bit error rate tester. At 77 K, the eye was closed with the nominal power voltage 3.3 V. The device resumed working when the voltage was increased to 3.6 V and continued to work up to 4.0 V, the absolute maximum rating of the supply voltage. We did not test with the supply voltage beyond 4.0 V. The eye diagrams at RT and at 77 K and 3.6 V are shown in Figure 6. The data shown in Figure 6 are $2^7-1$ PRBS at 2.5 Gbps. The jitter and amplitude at 77 K and 3.6 V and at RT and 3.3 V fall within the specifications. The large eye opening shown in the figure demonstrates the functionality of the laser driver at 77 K.

## 3. Cryogenic electrical links

For electrical data links, we tried to answer whether LVDS signals can be transmitted over 20-meters CAT5E twisted pairs at the liquid argon temperature. We tested a COTS LVDS driver (part number DS10BR150 from National Semiconductor) and a 20-meter CAT5E twisted pair at 77 K. Figure 7(a) is the block diagram of the test setup. Figure 7(b) is a picture of the test setup. The $2^7-1$ PRBS data were generated in a pattern generator (Model MP1763C from Anritsu) and sent to the LVDS repeater. The output signals were sent to a real time oscilloscope (Model DSA72001 from Tektronix) to measure eye diagrams or an error detector (Model MP1764C from Anritsu) to measure BER. At both ends of the CAT5E cable, we used a small printed circuit board (PCB) to connect the twisted pair to SMA connectors. The LVDS driver and the CAT5E cable were dipped in liquid nitrogen.

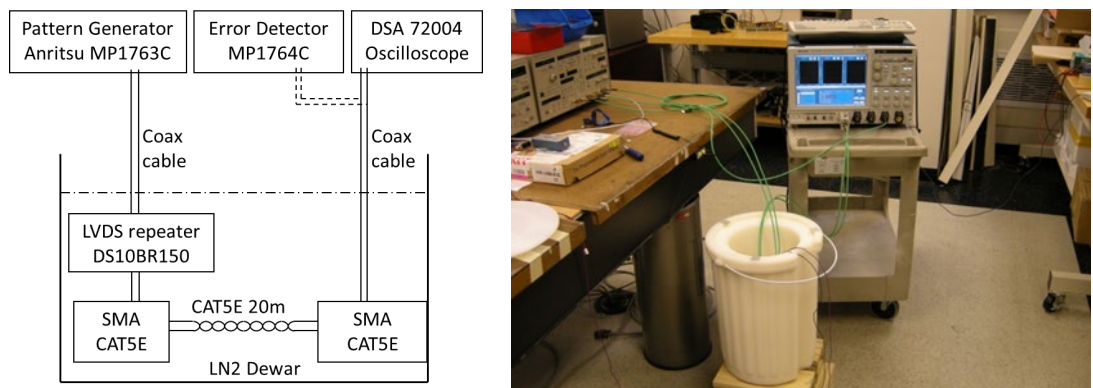

Figure 7: (from left to right) the block diagram (a) and a picture (b) of the test setup

The eye diagrams at 1 Gbps at RT and at 77 K are shown in Figure 8. The eye diagrams at 77 K have larger amplitude, faster edges, smaller jitter and noise, and wider eye opening than those at RT. In other words, the electrical link works better at 77 K than at RT. The BER of the electrical link at 77 K and 1 Gbps is $9.1\times10^{-13}$ at 90% confidence level, verifying that the LVDS driver can drive 20-meter CAT5E twisted pair at the data rate up to 1 Gbps at 77 K.

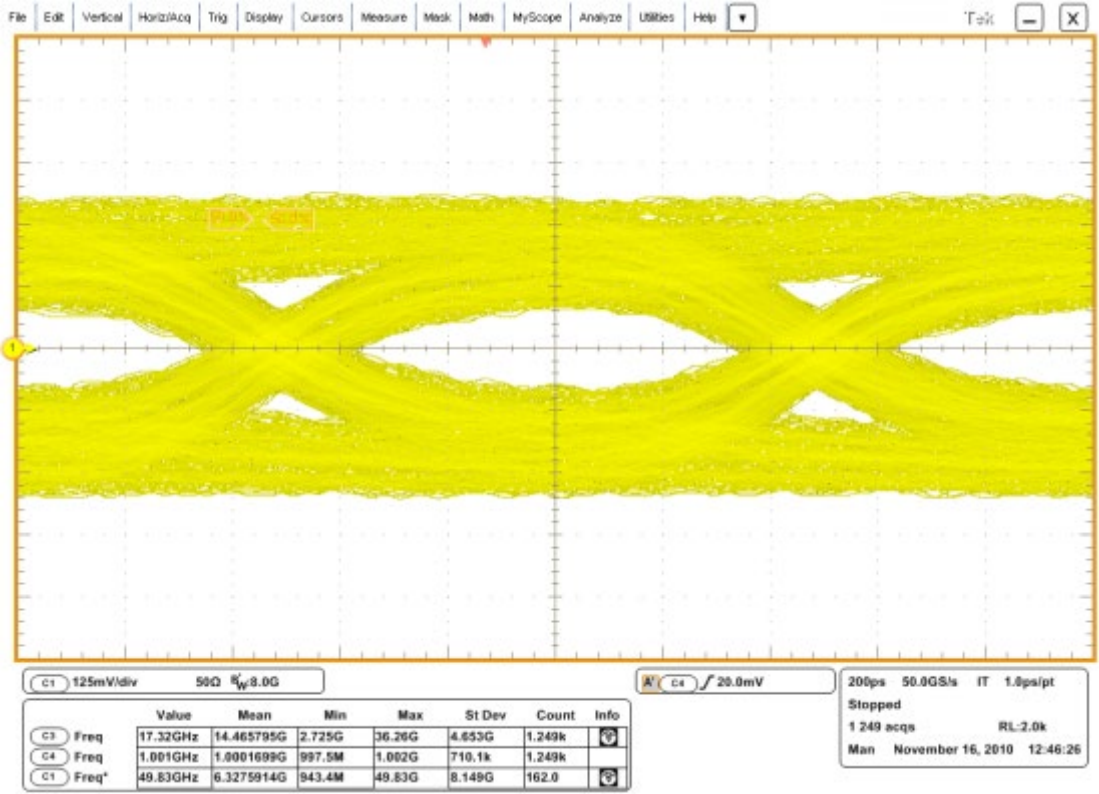
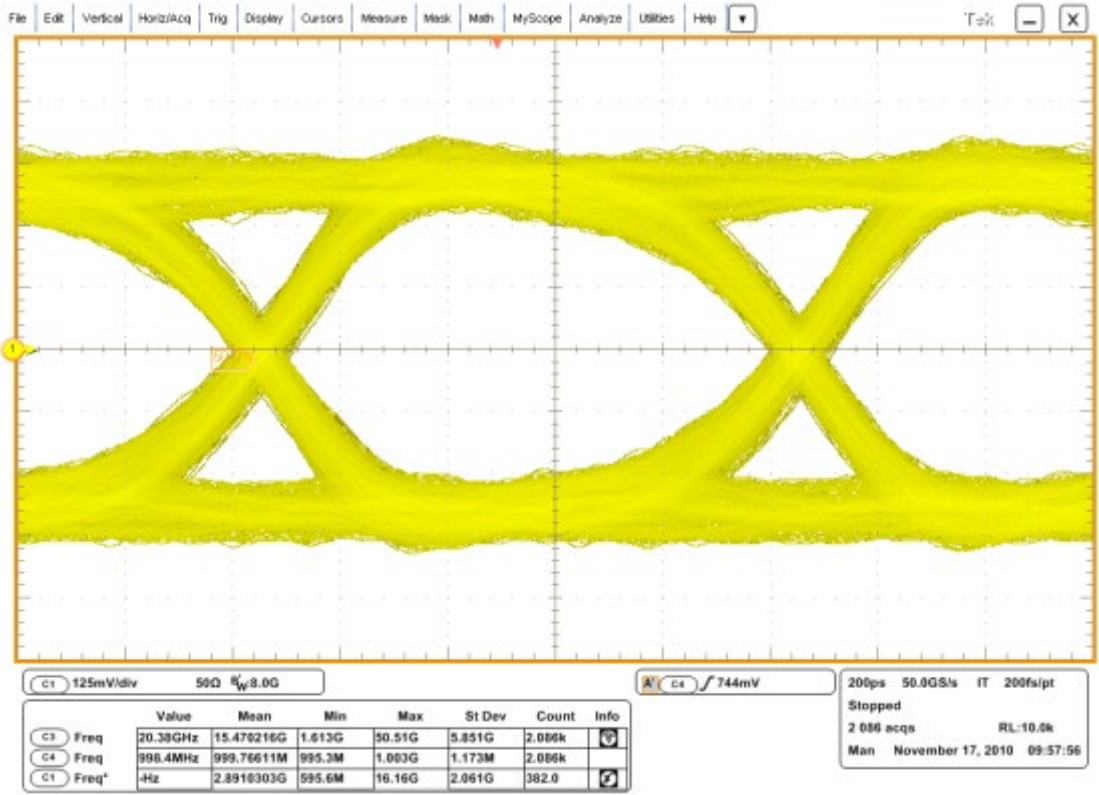

Figure 8: (from top to bottom) eye diagrams of the LVDS driver after 20-m CAT5E cables at 1 Gbps, RT (a) and 77 K (b)

## 4. Cryogenic FPGA

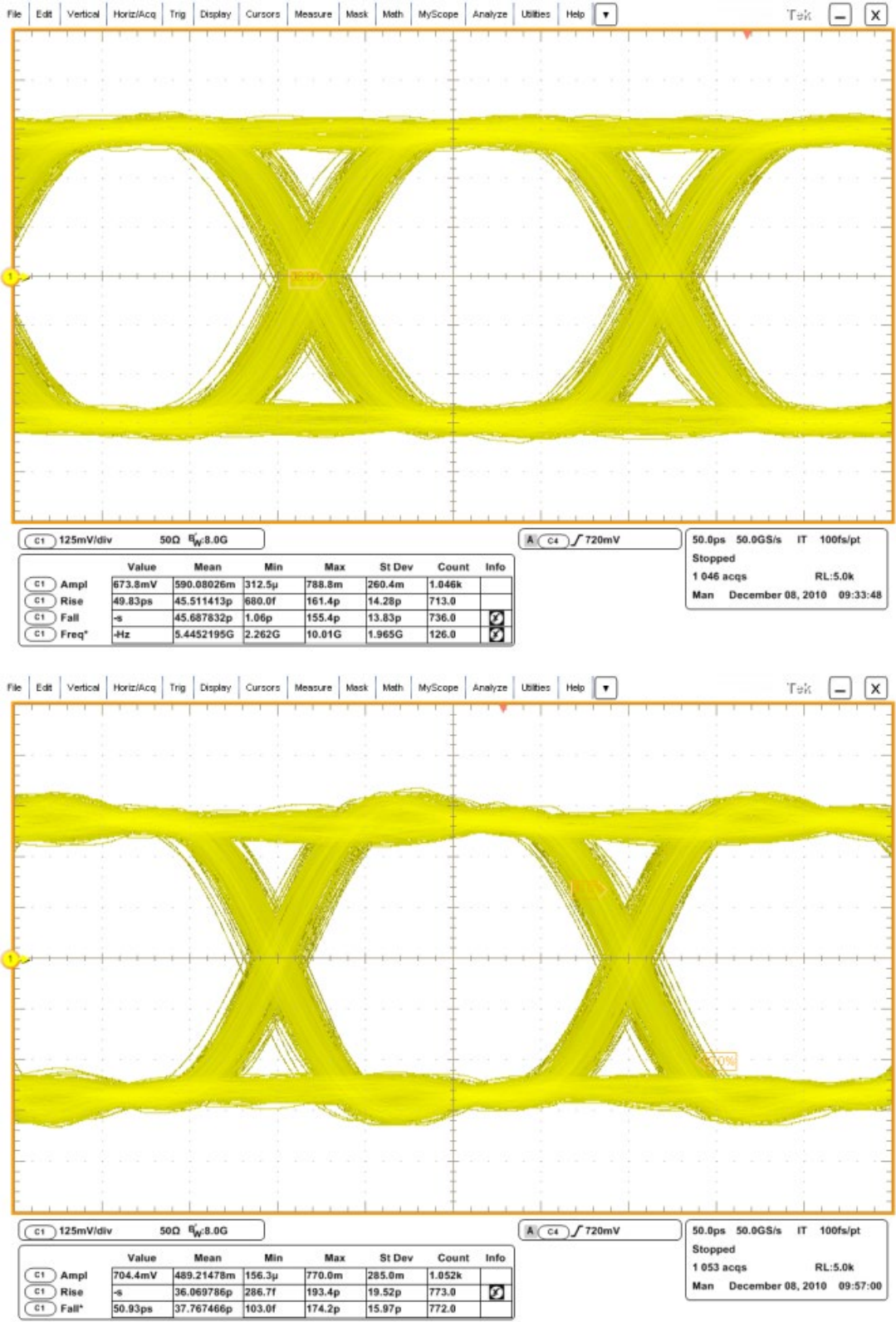

Figure 9: (from top to bottom) eye diagrams of EP2SGX90 at RT (a) and 77 K (b);

A Field Programmable Gate Array (FPGA) can be used for both electrical data links and optical data links. For electrical data links, the digital multiplexer and the cable driver can be implemented in an FPGA. For optical data links, the low-speed digital multiplexers, and encoder, and the serializer can be implemented in an FPGA.

An FPGA, EP2SGX90 in Stratix II GX series from Altera was tested at 77 K. With a clock input at 156 MHz, a $2^{23}$-1 PRBS generator is implemented at 5 Gbps inside the FPGA. The eye diagrams at RT (a) and at 77 K (b) are shown in Figure 9. The jitter, amplitude, and transition times are comparable at 77 K and at RT. Both BER's at RT and at 77 K are $3.8 \times 10^{-13}$ at 90% confidence level. The FPGA continues to function at 77 K.

Several FPGAs in Cyclone II series from Altera were tested at 77 K. A PRBS generator, a ring oscillator based on 17 stages of NOT gates, and a phase clocked loop (PLL) were implemented. For the PRBS generator, eye diagrams were measured. The BER was measured with no error during the test. For the ring oscillator, the differential amplitude, rise time and fall time, and power supply current were measured. For the PLL, the tuning range and jitter were measured. All measured parameters fell into the ranges specified in the data sheet.

The configuration memory EPCS4 from Altera was tested at 77 K. FPGAs can be programmed using configuration memory correctly at 77 K.

## 5. Cryogenic resistors and capacitors

Passive devices like resistors and capacitors are essential components for cold front-end electronics. For example, the decoupling and AC coupling capacitors from the wires to pre-amplifiers must be in liquid argon. The operation of resistors and capacitors in cryogenic temperature has been reported in literatures with a lot of non-consistence [19-22]. To study the performance of passive components, we purchased all types of resistors and capacitors available at the market and measured the parameters of components (1) at RT before they are dipped into liquid nitrogen, (2) at 77 K when they are dipped in the liquid, (3) at RT twenty minutes after they are taken out of liquid nitrogen, and (4) at 77 K when they dipped in the liquid again. At each condition, we measured at 100 Hz, 1 kHz, and 100 kHz with an LCR meter (Model 720 from Stanford Research System). For most resistors and capacitors (except electrolytic capacitors), the difference between frequencies is small.

The parameter changes at 1 kHz from RT to 77 K are listed in Table 1. The nominal value of each device is also listed. The number of samples in each type was usually only a few, but we tried to cover the parameter range as large as possible. If more than one sample in a type were measured, the parameter change was listed as a range. At 77 K, the resistance of the carbon composition resistors increase 19% and the resistance of metal element, wire wound, carbon film, thin film, metal film, and thick film change less than 7%. Capacitances of C0G ceramic, film and mica capacitors change less significantly than those of electrolytic capacitors and U/X/Y/Z ceramic capacitors. Resistance and capacitance in the two measurements at 77 K change about (only metal element resistors) or less than 1% (all other resistors and capacitors), meaning the measurements at 77 K are repeatable.

Table 1: Parameter change from RT to 77 K

| Type | Parameter change from RT to 77 K |
|---|---|
| Metal Element resistor (1 Ω) | -6.59% |
| Carbon Composition resistor (100 Ω) | 19.10% |
| Carbon Film resistor (10 k Ω) | 6.40% |
| Thin Film resistor (10 k Ω) | -0.21% |
| Metal Film resistor (10 k Ω) | 0.08% |
| Wire wound resistors (1 k, 10 k Ω) | 0.15% ~ 5.6% |
| Thick Film resistors (1, 100, 240, 1 k, 10 k, 100 k Ω) | 1.1% ~ 4.6% |
| Aluminum electrolytic capacitors | -100% |
| NbO electrolytic capacitors (4.7 μ, 10 μ, 100 μ F) | -71% ~ -39% |
| Tantalum Electrolytic capacitors (1 n, 10 n, 100 n, 22 μ F) | -96% ~ -10% |
| C0G/NP0 ceramic capacitors (18 p, 10 nF) | -4.1% ~ 0.35% |
| U, X, Y, Z ceramic capacitors (10 nF) | -94% ~ -24% |
| Film capacitors (10 n, 16 n, 47 n, 0.1 μ, 0.47 μ F) | -13% ~ 3.8% |
| Mica capacitors (2.0 n and 2.2 n F) | -0.35% ~ -0.12% |

## 6. Summary and outlook

We have studied the cryogenic performances of an LVDS driver driving 20-meter CAT5E twisted-pair, a serializer, laser diodes, a laser driver, optical fibers, and optical connectors, FPGAs, resistors and capacitors. All tests conducted show that the cryogenic digital data links for the liquid argon time projection chamber are promising.

After the cryogenic functionality on component level, another important issue is the reliability. We are studying the reliability of the cold electronics components and will report it in the future.

## Acknowledgments

The authors would like to express the deepest appreciation to Bob Biard, Gary Evans, Jim Guenter, Todd Huffman, Jim Tatum, Jan Troska, Francois Vasey, and Anthony Weidberg for beneficial discussions. The authors would like to thank Pritha Khurana, Arthur Mantie, Mark Schuckert and Tengyun Chen who provided the laser diodes. The authors would also like to thank Alan Humason for liquid nitrogen purchasing.